\documentclass[pra, reprint, twocolumn, superscriptaddress, amsfonts, amssymb, amsmath]{revtex4-2}
\usepackage[english]{babel}

\usepackage{graphicx}

\graphicspath{{./Images/},{./ImagesAppendix/},
{./images/},{./imagesAppendix/}}

\usepackage{physics}
\usepackage{bm}
\usepackage{braket}
\usepackage{pgfplots}
\usepackage{array}
\usepackage{makecell}
\usepackage{amsmath}
\usepackage{tikz}
\usetikzlibrary{positioning, arrows.meta}
\usepackage{subfigure}
\usepackage{commath}
\usepackage{graphicx,bm}
\usepackage{verbatim}
\usepackage{flafter}
\usepackage{float}
\usepackage{varioref}

\usepackage{color}
\usepackage[colorlinks,citecolor=darkBlue,linkcolor=darkBlue,
urlcolor=blue,hyperindex]{hyperref}
\definecolor{darkBlue}{rgb}{0.08, 0.13, 0.4}

\definecolor{THc}{rgb}{0.9,0.3,0.2}
 
\begin{document}


\title{Controlled flow of excitations in a ring-shaped network of Rydberg atoms}
\author{Francesco Perciavalle}
\affiliation{Quantum Research Centre, Technology Innovation Institute, Abu Dhabi, UAE}
\affiliation{Dipartimento di Fisica dell’Universit\`a di Pisa and INFN, Largo Pontecorvo 3, I-56127 Pisa, Italy}
\author{Davide Rossini}
\affiliation{Dipartimento di Fisica dell’Universit\`a di Pisa and INFN, Largo Pontecorvo 3, I-56127 Pisa, Italy}
\author{Tobias Haug}
\affiliation{QOLS, Blackett Laboratory, Imperial College London SW7 2AZ, UK}
\author{Oliver Morsch}  
\affiliation{CNR-INO and Dipartimento di Fisica dell’Universit\`a di Pisa, Largo Pontecorvo 3, 56127 Pisa, Italy}
\author{Luigi Amico}
\thanks{On leave from the Dipartimento di Fisica e Astronomia ``Ettore Majorana'', University of Catania.}
\affiliation{Quantum Research Centre, Technology Innovation Institute, Abu Dhabi, UAE}
\affiliation{INFN-Sezione di Catania, Via S. Sofia 64, 95127 Catania, Italy}
\affiliation{Centre for Quantum Technologies, National University of Singapore 117543, Singapore}

\date{\today}

\begin{abstract}
Highly excited Rydberg atoms are a powerful platform for quantum simulation and information processing.
Here, we propose atomic ring networks to study chiral currents of Rydberg excitations. The currents are controlled by a phase pattern imprinted via a Raman scheme and can persist even in the presence of dephasing.  Depending on the interplay between the Rabi coupling of Rydberg states and the dipole-dipole atom interaction, the current shows markedly different features. The excitations propagate with a velocity displaying a characteristic peak in time, reflecting the chiral nature of the current. We find that the time-averaged current in a quench behaves similarly to the ground-state current. This analysis paves the way for the development of new methods to transport information in atomic networks.
\end{abstract}

\maketitle

\section{Introduction}

Ultracold Rydberg states are highly excited energy states
of atoms cooled down to microKelvin temperatures through optical or magnetic means~\cite{low2012experimental}. 
The long-range character of their large dipole-dipole interaction implies rich physical properties.
Besides the dipole blockade,  the dipole moments associated with the electronic transitions among Rydberg states can lead to fast resonant energy transfer over a characteristic long range distance~\cite{browaeys2020many}. 
As a result, Rydberg atoms trapped in engineered magneto-optic potentials provide  ideal platforms for implementing controllable quantum many-body
simulators~\cite{morsch2018dissipative, hinrichsen2000non, bluvstein2021controlling, browaeys2020many, barredo2015coherent, chen2023continuous}
and quantum information processors~\cite{saffman2010quantum, cong2022hardware}.  In particular, resonant dipole-dipole interactions between Rydberg states are particularly useful for describing the transport of excitations in atomic networks~\cite{browaeys2020many}. In this context, both incoherent~\cite{gunter2013observing,maxwell2013storage} and coherent~\cite{barredo2015coherent} excitation transfer in short chains of Rydberg atoms have been demonstrated. 

In this work, we study ring-shaped networks of Rydberg atoms. It is important to remark that Rydberg atoms can be arranged in the most varied geometries as shown in~\cite{schymik2020enhanced,barredo2016atom}. Thus, also rings are experimentally realizable. Our goal is to produce a chiral ``excitation current'' flowing along the ring. 
Controlled flow of excitations in spatially closed quantum networks is of particular interest in quantum technology. Effective chiral interactions in closed circuits of superconducting qubits have been demonstrated to  simulate quantum phases of matter with a new twist~\cite{roushan2017chiral,gong2021quantum}.
Realizing guided coherent flows in matter-wave networks lies at the basis of atomtronics~\cite{wright2013driving,eckel2014hysteresis,del2022imprinting,cai2022persistent,ryu2020quantum,krzyzanowska2022matter}. The defining goal of this emerging field in quantum technology is to conceive devices and sensors of practical interest for applied science and technology, as well as create current-based quantum simulators to study new features of many-body systems (see Refs.~\cite{amico2021roadmap,amico2022colloquium} for roadmaps and reviews). 
In the implementations considered so far, atomtronics operates with flows of neutral matter. Here, we carry out a new conceptual step in the field, in which a controllable  {\it flow occurs in terms of Rydberg excitations}, rather than of matter.  We show how this goal can be realized through the application of a suitable combination of laser fields implementing specific Raman transitions that effectively impart a phase structure to the Rydberg states~\cite{dalibard2011colloquium}. Because of the characteristic interaction occurring  between Rydberg atoms, the local phases result in a chiral excitation current analog to matter-wave  currents~\cite{wright2013driving}. In contrast to persistent currents, the excitations propagate with specific properties that depend on the interplay between the coupling between Rydberg states, dipole-dipole interaction, and system size. This opens up the potential of new types of sensors and information transfer in atomic systems.

The paper is structured as follows. In Sec.~\ref{sec:Model}, we introduce our physical system and sketch a  scheme to generate a non-trivial phase for the Rydberg interaction.  Then, we analyze the propagation dynamics of the excitations including the effects of dephasing (Sec.~\ref{sec:Excitations}) and also compare ground state and long time averaged currents (Sec.~\ref{sec:current}). Finally, in Sec.~\ref{sec:Conclusions}, we summarize our results 
and discuss some possible further developments of our study.

\section{Model \& Methods}
\label{sec:Model}

\subsection{Model}
\label{subsec:Model}

\begin{figure}[!t]
\centering
\includegraphics[width=1\columnwidth]{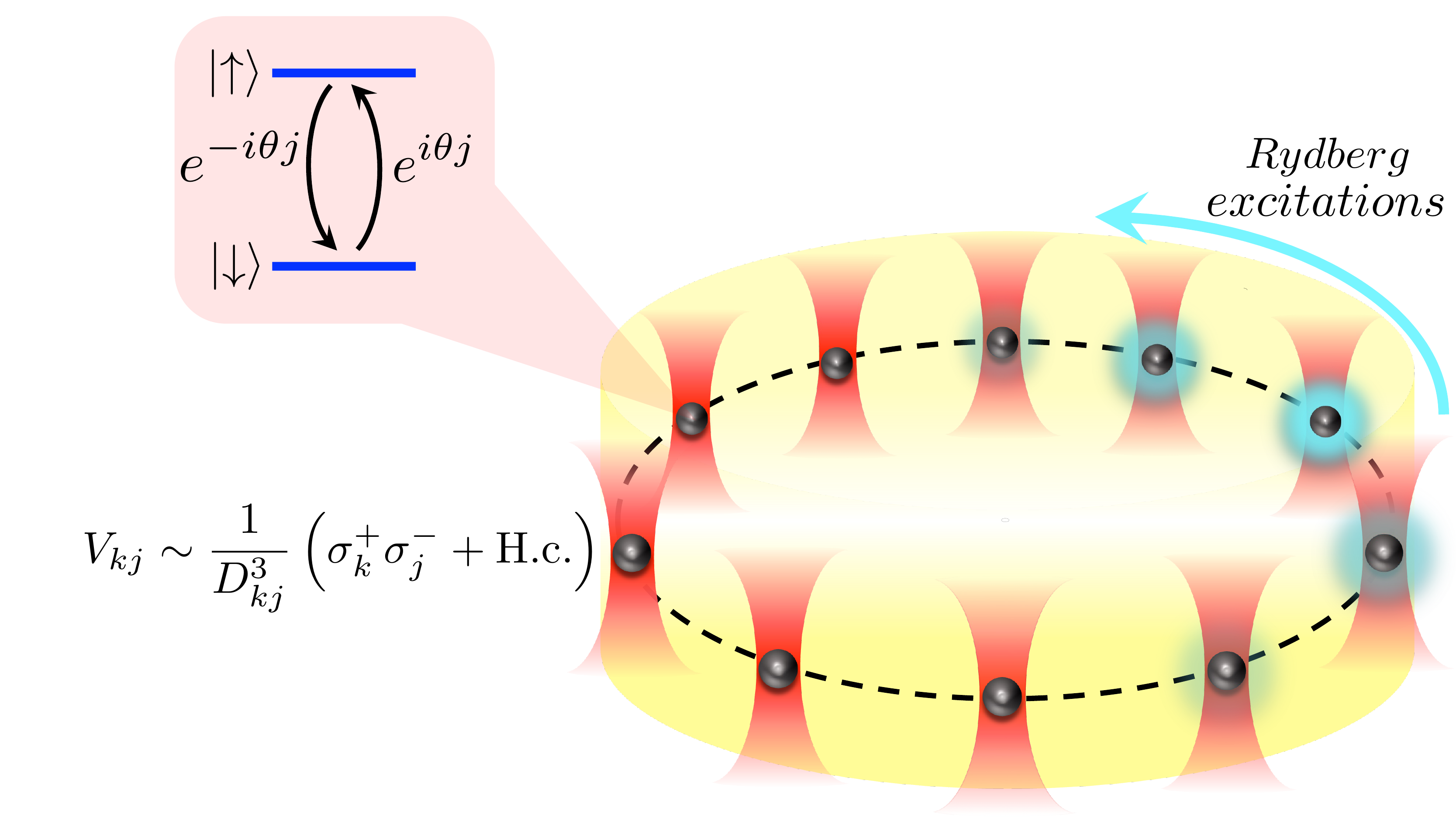}
\caption{The physical system: $N$ Rydberg atoms trapped by optical tweezers are arranged in a ring configuration. A spatially varying  phase factor on each atom is imparted via the local atom-laser interaction of the form $e^{i\theta j}\sigma_{j}^{+}+e^{-i\theta j}\sigma_{j}^{-}$ that couples the two Rydberg states $\ket{\downarrow_j}$ and $\ket{\uparrow_j}$. In the presence of dipole-dipole interactions, the phase imprinting yields a chiral transport of excitations.}
\label{fig:Sketch}
\end{figure}

We consider an array of $N$ Rydberg atoms arranged in a ring configuration (see Fig.~\ref{fig:Sketch}). The dipoles of a pair of Rydberg states of opposite parity $\ket{\downarrow_j}$, $\ket{\uparrow_j}$ and $\ket{\downarrow_k}$, $\ket{\uparrow_k}$ interact resonantly giving rise to the following Hamiltonian coupling~\cite{browaeys2020many}:
\begin{equation}
\mathcal{H}_{\rm int} = C_3 \sum_{k\neq j}\dfrac{1-3\cos^{2}\Theta_{kj}}{D_{kj}^{3}} \big( \sigma_{k}^{+}\sigma_{j}^{-} + {\rm H.c.} \big) ,
\label{eq:int}
\end{equation}
where $\Theta_{kj}$ is the angle between the quantization axis and the interatomic distance~\cite{barredo2015coherent,browaeys2020many}, while $D_{kj}=2R\sin(\pi|j-k|/N)$ is the distance between two atoms located in the ring at positions $j$ and $k$, $R$ being the radius of the ring. Here $\sigma^\alpha_j$  ($\alpha=x,y,z$) denote the spin-1/2 Pauli matrices on the $j$th atom and $\sigma^\pm_j = \tfrac12 (\sigma^x_j \pm i \sigma^y_j)$ are the corresponding raising/lowering operators, such that
$\sigma^{+}_j \ket{\downarrow_j}=\ket{\uparrow_j}$ and $\sigma^{-}_j \ket{\uparrow_j}=\ket{\downarrow_j}$.
We set the quantization axis in the direction orthogonal to the plane in which the atoms are located in such a way that $\cos(\Theta_{kj})=0$ $\forall k,j$ and the interaction is isotropic~\cite{chen2023continuous}. Then we consider a Rabi coupling term between the two Rydberg states with a spatially varying phase, so that the resulting full Hamiltonian can be cast as 
\begin{align}
\mathcal{H}=&\dfrac{\Omega}{2}\sum_{j} \big( e^{i\theta j}\sigma_{j}^{+} + {\rm H.c.} \big) +\dfrac{\Delta}{2}\sum_{j}\sigma_{j}^{z}+ \nonumber
\\& \sum_{k<j}J_{kj} \big( \sigma_{k}^{+}\sigma_{j}^{-} + {\rm H.c.} \big),
\label{eq:Ham_origin}
\end{align}
where $\theta=2\pi \ell/N$, $\Delta=\omega_{\uparrow\downarrow}-\omega_{L}$ is the detuning ($\omega_{\uparrow\downarrow}$ being the level spacing between the two Rydberg states), $\Omega$ is the Rabi frequency, and $J_{kj} = 2 C_{3}/D_{kj}^{3}$ describes the isotropic dipole-dipole interactions between atoms.

\begin{figure}[!t]
\centering
\includegraphics[width=1\columnwidth]{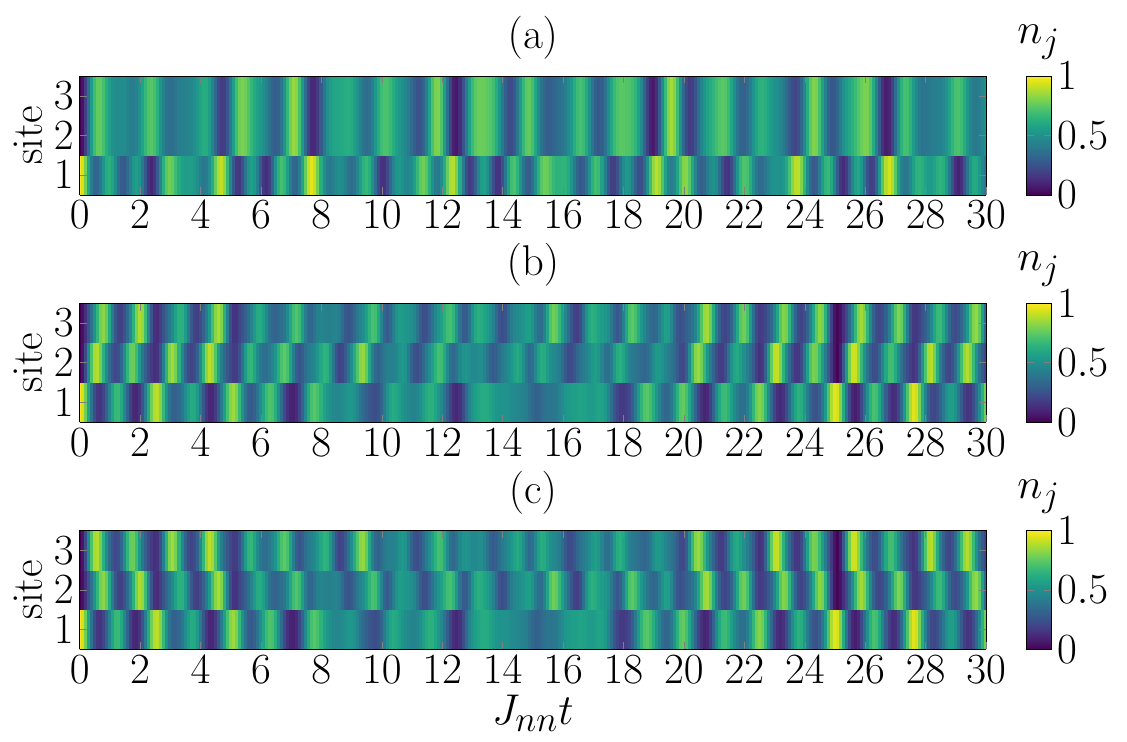}
\caption{Dynamics of the number of excitations $n_{j}=\left(1+\sigma_{j}^{z}\right)/2$ for three different values of the phase $\theta$: (a) $\theta=0$, (b) $\theta=2\pi/3$, (c) $\theta=4\pi/3$. $J_{nn}/\Omega=4$, and $N=3$. Data are obtained without dissipation.}
\label{fig:N3_dyn}
\end{figure}
\begin{figure*}[t]
\centering
\includegraphics[width=1\textwidth]{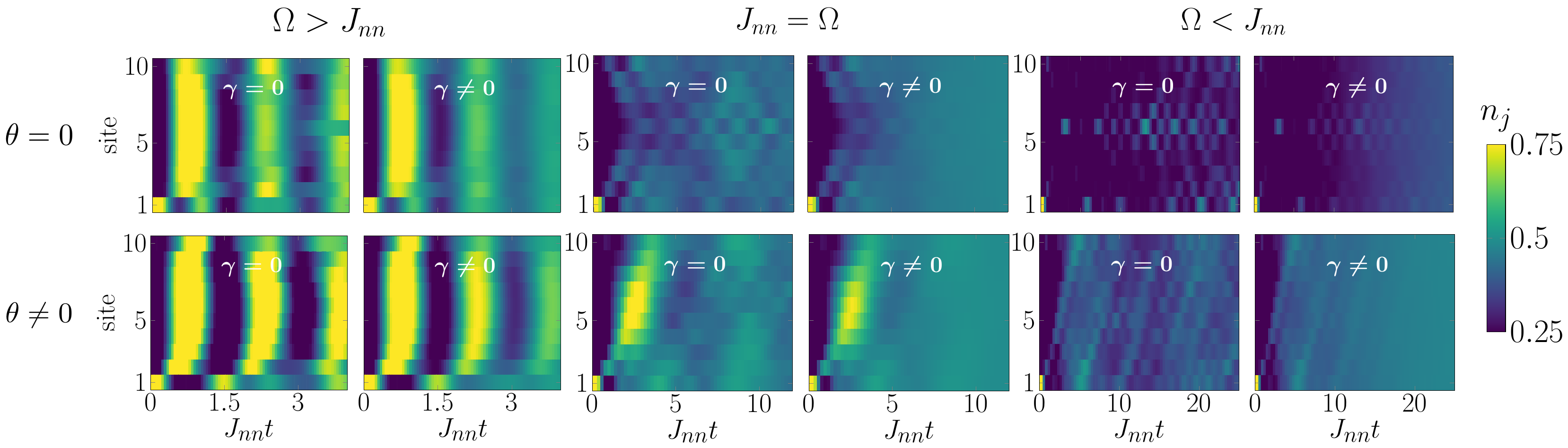}
\caption{Dynamics of Rydberg excitations.  The upper panels show the excitation propagation for $\theta=0$, while the lower panels display the excitation response to a phase winding $\ell=2$ ($\theta=2\pi/5$).  Results are shown for dephasing strength $\gamma=0$ and $\gamma\neq 0$ ($\Omega / \gamma = 20$).
  Left block of panels: $J_{nn}=1$, $\Omega=4$. Central block of panels: $J_{nn}=\Omega=1$.
  Right block of panels: $J_{nn}=1$, $\Omega=0.5$. Rings with $N=10$ atoms are considered.  }
\label{fig:N8_dynamics}
\end{figure*}

A possible implementation of the Hamiltonian in Eq~\eqref{eq:Ham_origin} makes use of  two Rydberg states of opposite parity, such as a pair of $S$- and  $P$-states.  In these conditions the atoms interact via~\eqref{eq:int}. In our scheme, we couple the two states via a Rabi term with a spatially varying phase $\theta j$ as depicted in Fig.~\ref{fig:Sketch} and we obtain the Hamiltonian~\eqref{eq:Ham_origin}. The spatially varying phase coupling can be achieved by a Raman scheme to couple the $S$ and $P$ Rydberg states in a three-photon Raman process.  In this process, an option to imprint the phase pattern would be to use a Laguerre-Gauss beam~\cite{matsumoto2008generation,lerner2012shaping,rubinsztein2016roadmap,mcgloin2003applications}; the $\Lambda$-scheme is completed by a Gaussian beam and a micro-wave field in the GHz range, via intermediate low-lying $P$ and Rydberg $D$ states (see Appendix \ref{appA}). For example, in a ring with $N=10$ atoms and an inter-atomic distance of a few $\mu m$, the waist of the Laguerre-Gauss beam can be of the order of  $10\mu m$, which is experimentally feasible. Alternative approaches to implement the site-dependent phase are suitable as well~\cite{wu2022manipulating,yang2022quantum}.

We now rotate the Hamiltonian~\eqref{eq:Ham_origin} using the following unitary transformation:
\begin{equation}
\mathcal{H} \rightarrow\mathcal{U}\mathcal{H}\mathcal{U}^{\dagger},
\qquad
\mathcal{U}=\prod_{k}e^{-i\theta k \sigma_{k}^{z}/2},
\end{equation}
describing a rotation along the $z$-axis, so that
\begin{equation}
\sigma_{j}^{\pm} \longrightarrow\sigma_{j}^{\pm}e^{\mp i\theta j},
\qquad
\sigma_{j}^{z}\longrightarrow\sigma_{j}^{z}.
\end{equation}
We finally obtain the rotated Hamiltonian 
\begin{equation}
\mathcal{H} = \frac{1}{2}
\sum_{j} \Big( \Omega\sigma_{j}^{x} + \Delta\sigma_{j}^{z} \Big)+ \sum_{k<j}J_{kj} \big( e^{i\theta(j-k)}\sigma_{k}^{+}\sigma_{j}^{-} + {\rm H.c.}  \big).
\label{eq:Ham_rotated}
\end{equation}

To address the transport of excitations, we study the equation of motion of the number of excitations $n_{j}=(1+\sigma_{j}^{z})/2$. We evolve the operator following the Heisenberg equation of motion $d n_{j} / dt = i\left[\mathcal{H},n_{j} \right]$ (in units of $\hbar=1$). Using the  commutation rules between Pauli matrices $\big[ \sigma^{\alpha}_j,\sigma^{\beta}_k \big]=2 i \, \varepsilon_{\alpha\beta\gamma} \, \delta_{jk} \, \sigma^{\gamma}_j$, we obtain
\begin{equation}
    \dfrac{dn_{j}}{dt}=-\nabla \mathcal{I}_{j}^{(nn)} + i\!\!\!\sum_{k\neq j,j\pm 1} \!\! J_{kj} \big( e^{i\theta( j- k)}\sigma_{k}^{+}\sigma_{j}^{-} - {\rm H.c.} \big) + \dfrac{\Omega}{2}\sigma_{j}^{y},
    \label{eq:eqmotion}
\end{equation}
where
\begin{eqnarray}
    \nabla \mathcal{I}_{j}^{(nn)} & = & \mathcal{I}_{j-1}^{(nn)}-\mathcal{I}_{j}^{(nn)},
    \\
    \mathcal{I}_{j}^{(nn)} & = & -i J_{nn}\left(e^{i\theta}\sigma_{j}^{+}\sigma_{j+1}^{-}-{\rm H.c.}\right).
\end{eqnarray}
We introduced the nearest-neighbor hopping $J_{nn}=2C_3/D_{nn}^3$, $D_{nn}=2R\sin(\pi/N)$ being the distance between two nearest neighbors atoms in the ring. The first two terms of Eq.~\eqref{eq:eqmotion} are related to the transport of excitations. In particular, the first term describes the transport between nearest-neighbor atoms, which is governed by the nearest-neighbor current scaling as $J_{nn} \sim 1/D_{nn}^{3}$. The second term describes the longer-range transport (beyond nearest-neighbor) and scales as $J_{kj} \sim 1/D_{kj}^{3}$, where $k$ and $j$ are at least next to nearest neighbors atoms. Due to the almost polynomial decay of $D_{kj}$ with the distance $\vert j-k\vert$, the nearest-neighbor current is the most important one. Both terms conserve the number of excitations. Finally, the last term of Eq.~\eqref{eq:eqmotion} describes the creation and destruction of excitations, due to the Rabi coupling $\Omega$. We note that role of the dipole-dipole interaction is crucial in our scheme; without that, the phase factors can be gauged away from the Hamiltonian and therefore they do not  produce any effect.

\subsection{Methods}
\label{subsec:Methods}
We initialize the system in the state $\ket{\psi(0)} = \ket{\uparrow \downarrow  \cdots \downarrow}$~\cite{barredo2015coherent} and evolve it with~\eqref{eq:Ham_rotated} to get $\ket{\psi(t)}=e^{-i {\cal H} t}|\psi(0)\rangle$, where we assume $\hbar = 1$. Then, we focus on the density of excitations
\begin{equation}
n_j(t) = \langle \psi(t)| \tfrac12 (1 + \sigma^z_j)|\psi(t) \rangle
\end{equation}
and study the flow of excitations through the nearest-neighbor current 
\begin{equation}
\mathcal{I}_{j}^{(nn)}(t) = -i J_{nn} \langle \psi(t)| \big( e^{i\theta}\sigma_{j}^{+}\sigma_{j+1}^{-} - {\rm H.c.} \big) |\psi(t) \rangle.
  \label{eq:currentNN}
\end{equation}
Due to the sensitivity of the Rydberg states to noise in the laser fields and fluctuations of the atom positions, decoherence can be present in the system~\cite{morsch2018dissipative}. In the following, we probe the robustness of the scheme by dephasing. We note that other sources of decoherence, like relaxation of the Rydberg states to the ground state, are expected to occur on much longer timescales than the excitation transport ones, and ultimately depending on the specific experimental implementation.
In practice, we consider a Markovian master equation for the system's density
matrix $\rho$~\cite{BreuerPetruccione}: \color{black} 
\begin{equation}
  \dfrac{\partial \rho}{\partial t}= -i[\mathcal{H},\rho] + \gamma \sum_{j} \mathcal{D}_\rho [\sigma_j^{z}] ,
  \label{eq:Master_eq}
\end{equation}
with $\mathcal{D}_\rho[O] \!=\! O \rho O^\dagger \!-\! \tfrac12 \{ O^\dagger O, \rho \}$
and $\gamma$ the dephasing rate. 

Finally, we compare a suitable time-averaged current with that of the ground state  $\mathcal{I}_{GS}^{(nn)}=\braket{\psi_{GS}|\mathcal{I}^{(nn)}|\psi_{GS}}$ which, in case of only nearest-neighbor hopping, corresponds to the derivative of the ground-state energy with respect to $\theta$. The ground state can be prepared in experiments using adiabatic protocols~\cite{bernien2017probing}.
In our calculations, we employed a combination of numerical methods, including exact diagonalization and density-matrix renormalization group approaches, to simulate the full long-range-interacting Hamiltonian~\eqref{eq:Ham_rotated}. The presence of the phase in the hopping gives rise to a non-zero chiral current which results in a directional transport of excitations. However, because of the Rabi term, the number of excitations is not conserved, and therefore, in spite of the phase imprint, the Rydberg excitation flow can be deteriorated by Rabi pumping. We shall see that a directionality does emerge in the dynamics of the number of excitations and can be optimized by adjusting  $\Omega$ and the interaction strength. In addition, the presence of the Rabi coupling and its interplay with atom interactions induces excitation flows with different features.

We note that a chiral flow of Rydberg excitations has been observed in a system of $3$ atoms with triangular geometry~\cite{lienhard2020realization}.  With this approach the phase is imprinted effectively by a suitable tuning of the amplitude of external fields. In particular, depending  on the geometry of the system and the number of atoms,  the phase factors and hopping amplitudes cannot be controlled independently. For the  hexagonal lattice system considered in Ref.~\cite{lienhard2020realization}, these constraints hinder the chirality of the flow.  We also note that our proposal is substantially different from protocols relying on spin-orbit coupling in cold atom systems in  which the flow occurs in terms of matterwaves~\cite{celi2014synthetic,mancini2015observation,stuhl2015visualizing}.

\section{Excitation dynamics}
\label{sec:Excitations}

We first address the case with $N=3$ atoms, where, due to the small size, we only have a nearest-neighbor hopping. This case is instructive to understand the role of non-zero phase in the system considered. In Fig.~\ref{fig:N3_dyn} we plot the dynamics of the number of excitations $n_j(t)$ for three different values of the phase: $\theta=0$, $\theta=2\pi/3$ and $\theta=4\pi/3$; we assume zero detuning ($\Delta=0$). The system is initialized in a configuration with a single localized excitation: $\ket{\psi(0)}=\ket{\uparrow_1, \downarrow_2, \downarrow_3}$. In the absence of a phase ($\theta=0$), the dynamics of $n_{2}$ and $n_{3}$ is symmetric and there is no a preferred direction in the excitation transport. The presence of a non-zero phase has a visible effect on the dynamics, thus providing directionality to the system through the breaking of symmetry between $n_{2}$ and $n_{3}$. In particular, for $\theta=2\pi/3$ the excitation moves clockwise, following the path ``$1\rightarrow 2 \rightarrow 3 \rightarrow 1 ...$". 
The presence of the Rabi pumping in the Hamiltonian can deteriorate the directional transport leading to suppression and revival of the transport. We also show the dynamics of excitations for $\theta=4\pi/3$ (which corresponds to $\theta=-2\pi/3$), where the excitations flow in the opposite (anti-clockwise) direction, compared to that for $\theta=2\pi/3$. 

\begin{figure}[!t]
\includegraphics[width=\columnwidth]{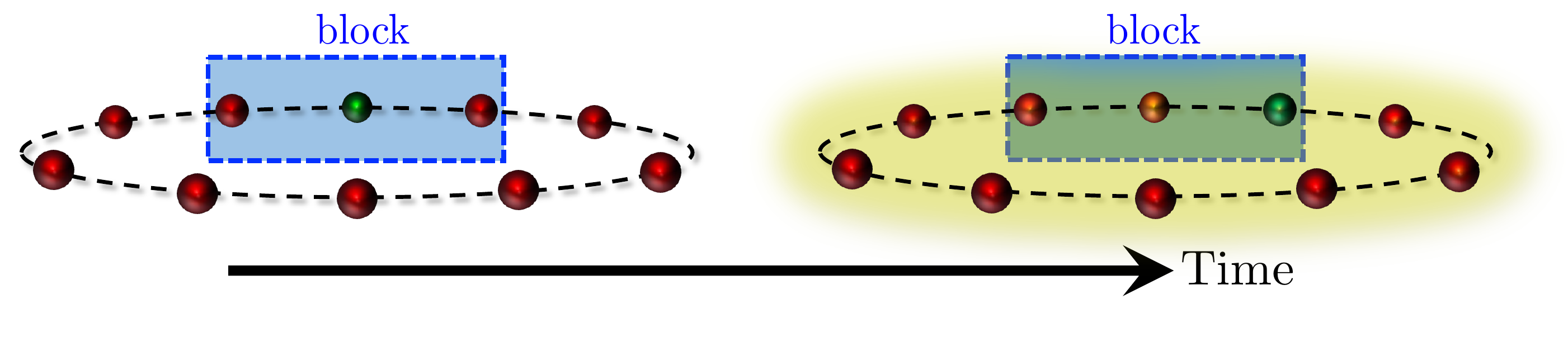}
\caption{Sketch of the adopted protocol: The system is initialized with a single excitation localized in the central site of a three-atom block and then allowed to evolve freely in time; mean excitation position and its velocity in the block are used as figures of merit to study the chirality of the flow once a phase pattern is imprinted to the ring (represented by a yellow shine in the figure).}
\label{fig:scheme-XV}
\end{figure}

Results for the dynamics of $n_j(t)$ with $N=10$ are summarized in Fig.~\ref{fig:N8_dynamics}.
Also in this case, for $\theta=0$, the excitations are found to spread reciprocally to the left and to the right with an equal probability. In contrast, for finite values of $\theta$, the flow can be directional with a $2\pi$-periodicity (for $\theta\rightarrow -\theta$ it changes direction). 
A nontrivial dependence on the interplay between hopping and $\Omega$ can be observed. In particular, for $\Omega>J_{nn}$ the local density of excitations at each site oscillates in time. The flow is directional only on short intervals around intermittent times, while in between it is suppressed by the effect of the Rabi interaction $\Omega$. Note the emergence of a collective character involving different sites: after moving from the initial to the nearest-neighbor sites, the excitation spreads through the entire ring, except the starting site and its nearest neighbors; then it comes back to the site in which it was initially created.
In the $J_{nn}=\Omega$ regime, we clearly see a directionality of the excitation current propagating from one site to the next-neighbor one. In this regime, though, the directional flow is suppressed at longer times.
For $\Omega<J_{nn}$, the flow is directional and persists at any time, however the fraction of excitation that moves in the ring is smaller compared to the two other regimes. 

In all the three explored regimes, a small dephasing is found not to be detrimental for the excitation flow. On the other hand, when it becomes comparable with the energy scales of the system the excitation flow can be suppressed (see also Appendices \ref{appA} and \ref{appB}).  We also find that the presence of a small detuning $\Delta$ does not affect the excitation flow (see Appendix \ref{appA}).

\subsection{Excitations velocity}%
\label{sec:Velocity}

We now analyze the velocity at which Rydberg excitations travel along the ring. For a quantitative study, minimizing spurious effects coming from the closed geometry and long-range hoppings, we load the excitation on a single site (say $j=0$), let the system freely evolve in time, and then focus on a suitable finite-site block around $j=0$ [see Fig.~\ref{fig:scheme-XV}]. The mean excitation position 
$x_{\operatorname{m}}^{(\rm block)}$ and its corresponding velocity $v_{\rm m}^{(\rm block)}$ are defined as 
\begin{equation}
x_{\operatorname{m}}^{(\rm block)} = \frac{\sum_{j=-1}^1 (D_{nn} \, j) n_{j}}{\sum_{j=-1}^{1} n_{j}}, \quad
v_{\rm m}^{(\rm block)} = \frac{d x_{\operatorname{m}}^{(\rm block)}}{dt},
\label{eq:exc_velocity}
\end{equation}

\begin{figure}[ht]
\centering
\includegraphics[width=0.85\columnwidth]{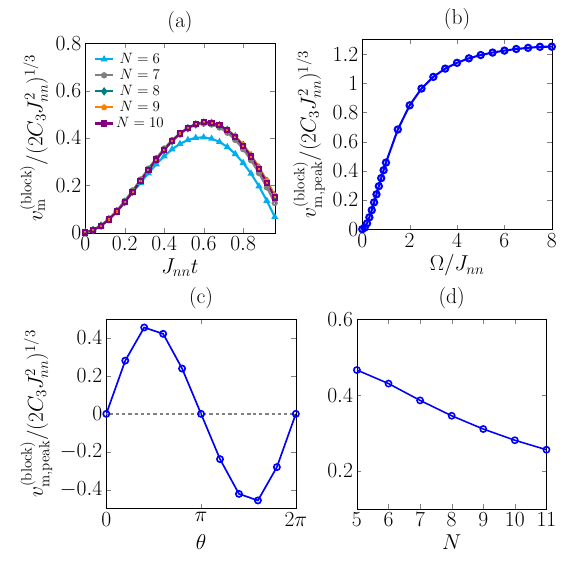}
\caption{Propagation velocity of the Rydberg excitations. (a) Dynamics of the dimensionless velocity for different numbers $N$ of atoms ($\Omega / J_{nn}=1$, $\theta=\pi/2$). (b)-(c) Dependence of the maximum  value reached $v^{(\rm block)}_{\rm m, peak}$, for $N=10$, on the ratio between the Rabi frequency $\Omega$ and the nearest neighbor hopping $J_{nn}$ (panel c: $\theta=2\pi/5$) and on the phase $\theta$ (panel d: $\Omega / J_{nn} = 1$). (d) Behavior of the dimensionless velocity as a function of the number $N$ of atoms ($\Omega / J_{nn}=1$, $\ell=1$). Measures are obtained in the absence of dephasing.}
\label{fig:CM_main}
\end{figure}

To clarify our physical understanding, we restrict our analysis to a block of atoms in the neighborhood of the central site (see Fig.~\ref{fig:scheme-XV}). By considering a small block around $j=0$, we avoid the $j=-N/2$ to $j=N/2$ discontinuity in the ring. Furthermore, due to the $1/D^3$ decay of interaction strength with distance $D$, long-range interactions are not dominant here.
The variations of $x_{\rm m}^{(\rm block)}$ and $v_{\rm m}^{(\rm block)}$ at short times provide information about the chirality, the velocity of excitations, and the amount of excitations that are moving. In fact, $x_{\rm m}^{(\rm block)}=0$ indicates that excitations move along the positive and the negative direction symmetrically, and thus there is no directionality. 
The sign of $x_{\rm m}^{(\rm block)}$ and $v_{\rm m}^{(\rm block)}$ indicates the direction in which excitations are moving. The modulus $| v_{\rm m}^{(\rm block)}|$ provides information on the typical velocity of excitations. We report the behavior of the dimensionless velocity computing the position in $D_{nn}$ units and the time in $1/J_{nn}$ units:
\begin{equation}
    \dfrac{d(x_{\rm m}^{(\rm block)}/D_{nn})}{d(J_{nn}t)}=\dfrac{1}{(2C_{3}J_{nn}^{2})^{1/3}}v_{\rm m}^{(\rm block)}. 
\end{equation}
At short times, these block-based quantities provide information on the dynamics of the fraction of excitations from the site $j=0$ to its nearest neighbors. Their sign and modulus are experimentally measurable quantities, whose behavior can be exploited to study the properties of the current and the amount of excitation moving in the ring. 

In Fig.~\ref{fig:CM_main} we focus our attention on the behaviour of the mean velocity of excitations as a function of the Hamiltonian parameters. First, in Fig.~\ref{fig:CM_main}(a), we show that $v_{\operatorname{m}}^{(\rm block)}$ is peaked in time independently on the size of the system. More details on the dynamics of the mean excitation position and velocity are provided in Appendix~\ref{app:EXCPOS}. In Fig.~\ref{fig:CM_main}(b,c,d) the peak velocity $v_{\operatorname{m, peak}}^{(\rm block)}$ is analyzed. The dependence on the Rabi frequency is displayed in Fig.~\ref{fig:CM_main}(b): larger values of $\Omega$ move a bigger amount of excitations on shorter time-scales and thus the peak velocity is bigger. In Fig.~\ref{fig:CM_main}(c) we report its shape with the phase $\theta$: we observe that $v_{\operatorname{m,peak}}^{(\rm block)}(\theta)=-v_{\operatorname{m,peak}}^{(\rm block)}(-\theta)$, due to the presence of a chiral current in the system. The behavior of $v_{\operatorname{m,peak}}^{(\rm block)}$ with the number of atoms $N$ is slowly decreasing, for fixed $\ell$ [Fig.~\ref{fig:CM_main}(d)]. For a system of $N$ Rydberg atoms with a spacing of a few $\mu m$, the typical nearest-neighbor coupling $J_{nn}$ is of the order of MHz~\cite{barredo2015coherent, chen2023continuous}, which corresponds to a velocity of the order of $\mu m/\mu s$.

\section{Time-averaged current}\label{sec:current}%
%
The system considered so far admits the presence of non-zero excitation currents; the study of their ground state and dynamical behaviour is instructive to understand the nature of the flow. In Appendix \ref{app:GScurr} we study the behaviour of the ground state current, showing how it has a chiral behaviour and features dependent on the ratio $\Omega/J_{nn}$. Here, we compare it with a long-time averaged one.  In particular, we initialize the system in the superposition state  $\ket{\psi_{sup}}=(\sqrt{N})^{-1}\sum_{j}\ket{\downarrow_{1}\cdots\uparrow_{j}\cdots\downarrow_{N}}$, evolve it under the Hamiltonian~\eqref{eq:Ham_rotated} and compute the long-time-averaged current
\begin{equation}
\bar{\mathcal{I}}^{(nn)}=\lim_{T\rightarrow\infty}\dfrac{1}{T}\int_{0}^{T}dt \braket{\psi(t)|\mathcal{I}^{(nn)}|\psi(t)},
  \label{eq:Long_time_average}
\end{equation}
where $\mathcal{I}^{(nn)}=N^{-1}\sum_j \mathcal{I}^{(nn)}_j$. We focus our attention on the regime $\Omega>J_{nn}$ and only on the dominant nearest-neighbor current; results of long range currents are shown in Appendix \ref{app:LR_currents}. The specific choice of an initial state as the spin-wave state $\ket{\psi_{sup}}$ is expected to generate a dynamics of excitations similar to a persistent current. 

\begin{figure}[H]
\includegraphics[width=\columnwidth]{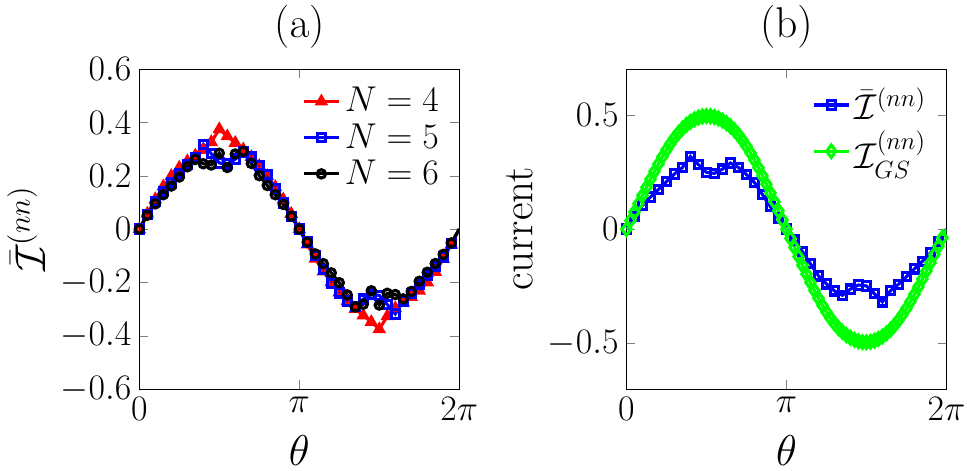}
\caption{Time-averaged vs ground-state current.
To compute $\mathcal{I}^{(nn)}$, we initialize the system in $\ket{\psi_{sup}}$. (a) The $N$-dependence of the long-time nearest-neighbor averaged current, for $\Omega=8$,  $J_{nn}=1$. (b) Comparison of $\mathcal{I}^{(nn)}$ with $\mathcal{I}_{GS}^{(nn)}$, for $N=5$, $J_{nn}=1$, and $\Omega=8$.}
\label{fig:Dyn_GS}
\end{figure}

Figure~\ref{fig:Dyn_GS} displays  $\bar{\mathcal{I}}^{(nn)}$  vs the ground-state persistent current $\mathcal{I}^{(nn)}_{GS}$. While differing in quantitative details, they share similar patterns and the same periodicity. Remarkably, the excitation current displays weak dependence on the system size. This behavior should be contrasted with the mesoscopic nature characterizing the matter-wave persistent current~\cite{imry2002introduction}. Here, the time-averaged current is computed in the absence of dephasing, which would eventually deteriorate the signal at long times. To obtain time averages, one would thus need to 
calculate Eq.~\eqref{eq:Long_time_average} not in the $T\rightarrow \infty$ limit, but up to a finite time, until convergence is reached.

\section{Discussion and Conclusions}
\label{sec:Conclusions}
%
We have carried out the first conceptual steps of  Rydberg atomtronics, i.e., cold-atom networks in which a directional current flows in terms of Rydberg excitations. Here, we demonstrate a chiral current in a ring-shaped spatial configuration. We propose a possible experimental scheme in which two Rydberg states are coupled by a suitable combination of laser fields (micro-wave, Gauss, and Laguerre-Gauss laser fields) such that a spatially varying phase on the excitation transfer is imprinted locally on each atom. Because of the Rydberg  dipole-dipole  interaction, such local phases produce a phase gradient in the many-body system which results in  a chiral current of Rydberg excitations. In our scheme, the flow can be controlled by the phase imparted by the Laguerre-Gauss field.  

The dynamical features of the excitation flow depend on the interplay between the dipole-dipole interaction $J_{kj}$ and the Rabi coupling $\Omega$ between the Rydberg states (see Fig.~\ref{fig:N8_dynamics}). In the regime $\Omega>J_{nn}$, the flow is clearly directional only for short intermittent time intervals and the excitations move collectively. For $\Omega<J_{nn}$, the flow is directional at any time, but with a lower density, indeed the  excitations move in fractions as in a `relay race': while a group of excitations travels, at  characteristic times another group starts. For $J_{nn}\simeq\Omega$, a substantial flow is obtained: in this regime, the excitations move from one site to the nearest-neighbor one. 
The effect of dephasing is notable: while at short times the excitation flow is robust, at long times it is hindered by dephasing and the steady state tends to be completely mixed.

The velocity of this process can be investigated at short times, initializing the system with a single localized excitation and monitoring the mean position of the excitations around its neighborhood (see Fig.~\ref{fig:scheme-XV}). Its velocity is non-zero and peaked in time.  {\it The change in time} of the mean position of excitations reflects the chiral nature of the current. 
The time-averaged current shows a pattern similar to that of the ground-state persistent current (see Fig.~\ref{fig:Dyn_GS} and~\ref{fig:I_om}). In particular, because of the Rabi coupling, both quantities are not mesoscopic.

Our study demonstrates how Rydberg atoms can realize a new concept of atomtronic networks based on a local engineering of the dipole-dipole interaction. In comparison with standard  implementations in which atoms move on millisecond time scales, the Rydberg platform has the potential to realize much faster devices operating on microsecond time scales. Owing to the specific coherent properties of the excitation flow demonstrated here, this scheme opens up the possibility for a controllable entanglement transfer along one-dimensional atom networks on a long-range spatial scale.  Other types of Rydberg atomtronic networks are left for future studies. Finally, the current of the Rydberg excitations can be readily measured via quantum gates (see Appendix~\ref{App:CurrentMeas}).
\begin{acknowledgements}
{\it Acknowledgments.} We thank Enrico Domanti, Juan Polo and Wayne J. Chetcuti for discussions. The Julian Schwinger Foundation grant JSF-18-12-0011 is acknowledged. OM also acknowledges support by the H2020 ITN ``MOQS" (grant agreement number 955479) and MUR (Ministero dell’Università e della Ricerca) through the PNRR MUR project PE0000023-NQSTI.
Numerical computations have been performed using the Julia packages \texttt{QuantumOptics.jl} and \texttt{Itensors.jl}~\cite{kramer2018quantumoptics,fishman2022itensor}.
\end{acknowledgements}

\appendix
\section{Possible experimental realization}
\label{appA}
\begin{figure}[H]
\centering
\includegraphics[width=0.8\columnwidth]{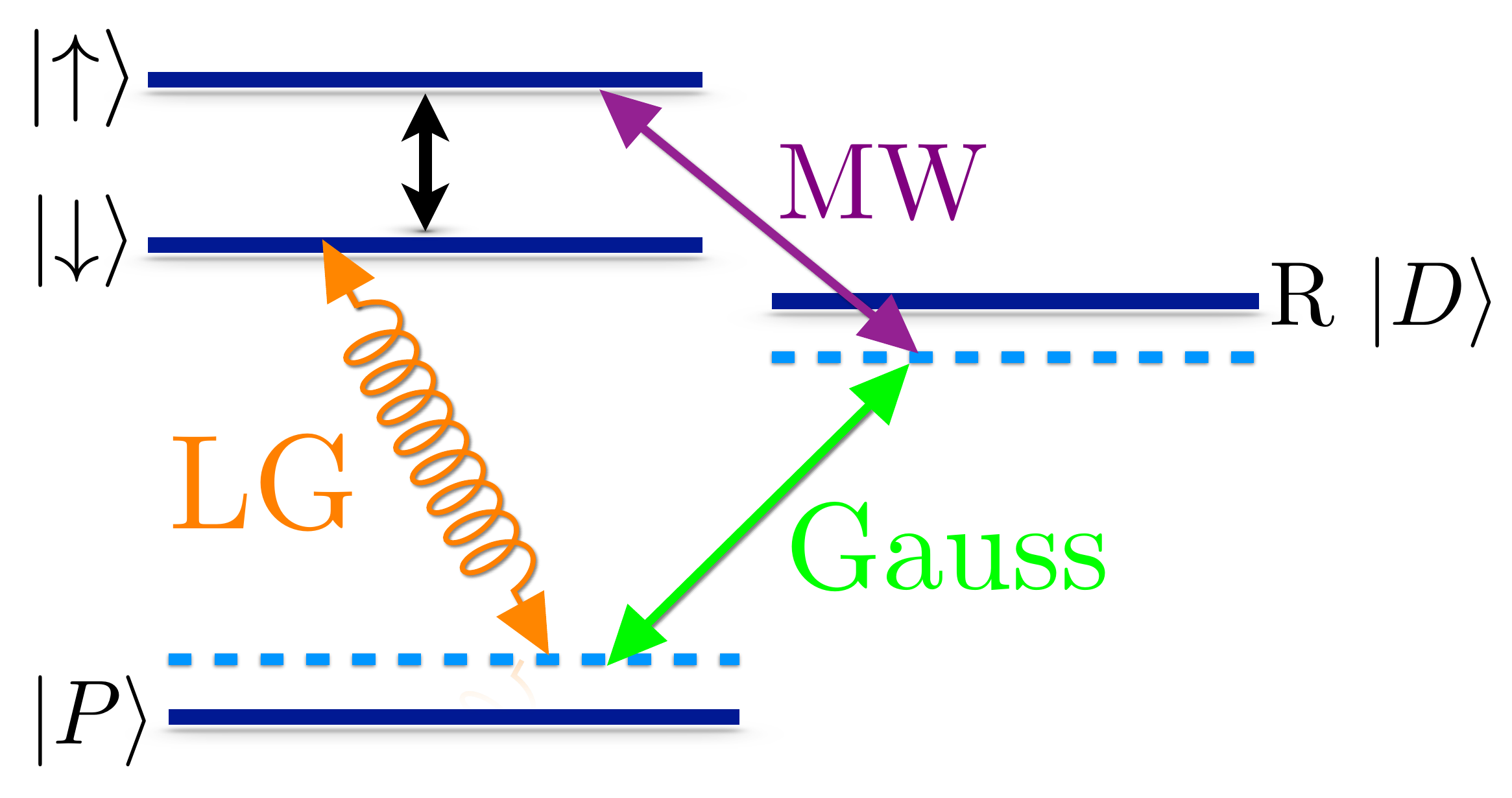}
\caption{Pictorial representation of a possible experimental scheme that can be used to realize a coherent coupling between the two Rydberg states with a spatially dependent phase.}
\label{fig:scheme_Ramann}
\end{figure}

We propose here a possible experimental scheme to couple coherently two Rydberg states $\ket{\uparrow}$ and $\ket{\downarrow}$ of opposite parity with a spatially varying phase (see Fig.~\ref{fig:scheme_Ramann}). A three-photon Raman process involving a Laguerre-Gauss beam (LG, orange), a Gaussian beam (Gauss, green) and a micro-wave (MW, purple) field in the GHz range is used to couple $S$ and $P$ Rydberg states (respectively $\ket{\downarrow}$ and $\ket{\uparrow}$) via intermediate low-lying $P$ and Rydberg $D$ states. The infrared and micro-wave fields are detuned sufficiently from the resonances with the intermediate states to ensure that those states are not populated. In this scheme, the phase pattern is imprinted using a Laguerre-Gauss beam. The phase is $\theta=2\pi\ell / N$, $\ell$ being the orbital angular momentum carried by the Laguerre Gauss field.

\section{Effect of non-zero dephasing and detuning on the dynamics}
\label{appB}
\begin{figure}[h]
\centering
\includegraphics[width=0.9\columnwidth]{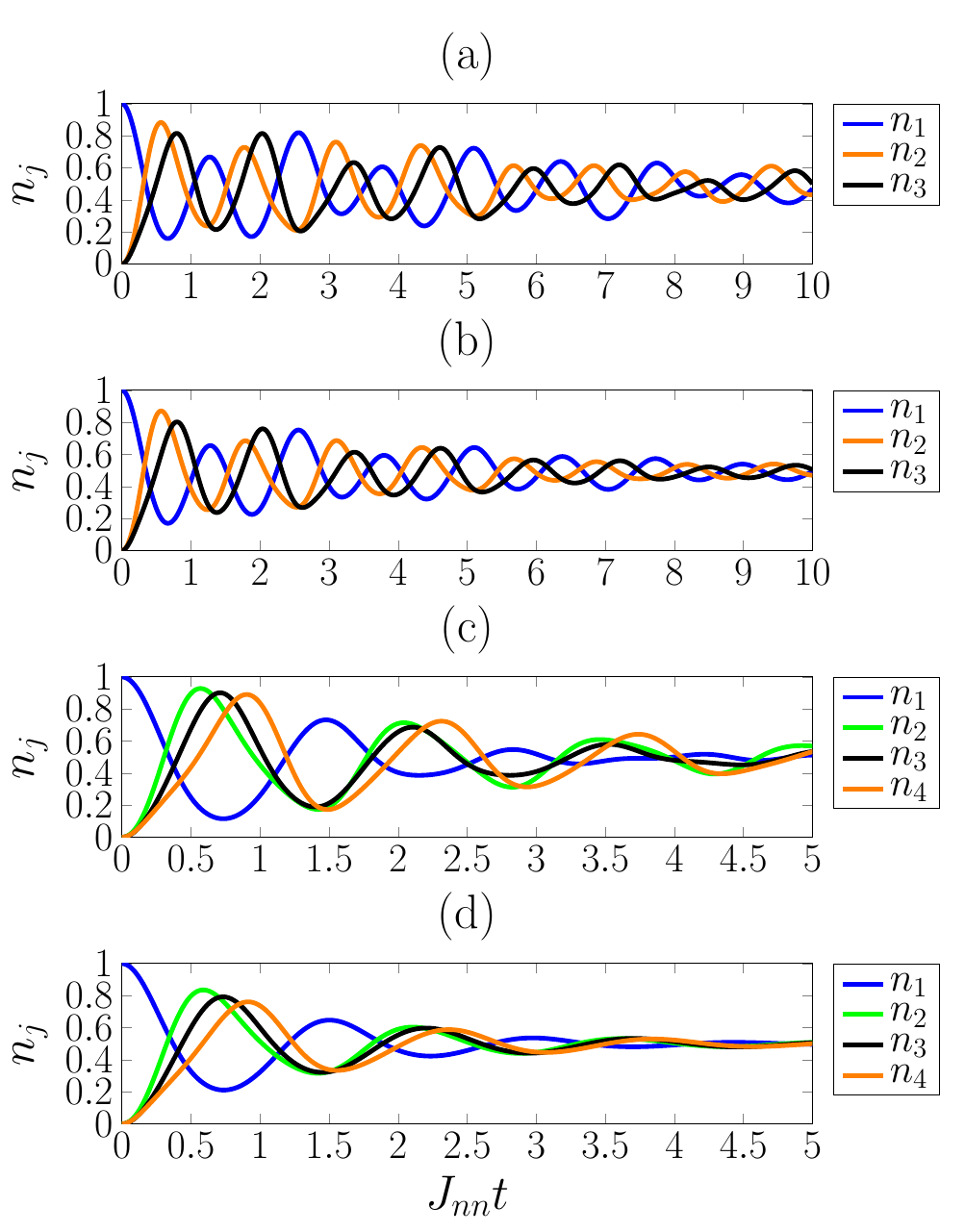}
\caption{Dynamics of the number of excitations $n_j=(1+\sigma_j^z)/2$ for small systems in presence of pure dephasing. In all the four cases $J_{nn}=1$ and $\Omega=4$ but different regimes of dephasing are reported. (a-b) $N=3$ and the phase is $\theta=2\pi/3$ ($\ell=1$); $\gamma=0.05$ (a) and $\gamma=0.1$ (b). (c-d) $N=4$ and the phase is $\pi/2$ ($\ell=1$); $\gamma=0.1$ (c) and $\gamma=0.5$ (d). The dephasing $\Delta$ is set to zero.}
\label{fig:N3_4_deph}
\end{figure}
We analyze in a more detailed way the effect of pure dephasing on the dynamics of the number of excitations for small systems ($N=3$ and $N=4$). In addition, we also study the effect of a non-zero detuning $\Delta$. In Fig.~\ref{fig:N3_4_deph}, we consider the effect of the dephasing on the dynamics of the number of excitations, ruled by Eq.~\eqref{eq:Master_eq}. We compare $N=3$ in which the nearest neighbor hopping does not come into play with $N=4$ in which it is present. From Fig.~\ref{fig:N3_4_deph}, we observe that directionality is present. In particular, for $N=3$ and in presence of dephasing, at all reported times the majority of excitations alternate regularly in a directional way ($1\rightarrow 2 \rightarrow 3 \rightarrow 1 ...$). However, a damping of the oscillation amplitudes inevitably sets in at longer times, thus suppressing oscillations and flow. At long times the system reaches a steady state in which the number of excitations in all the sites is $n_j^{SS}=1/2$.

\begin{figure}[!t]
\centering
\includegraphics[width=\columnwidth]{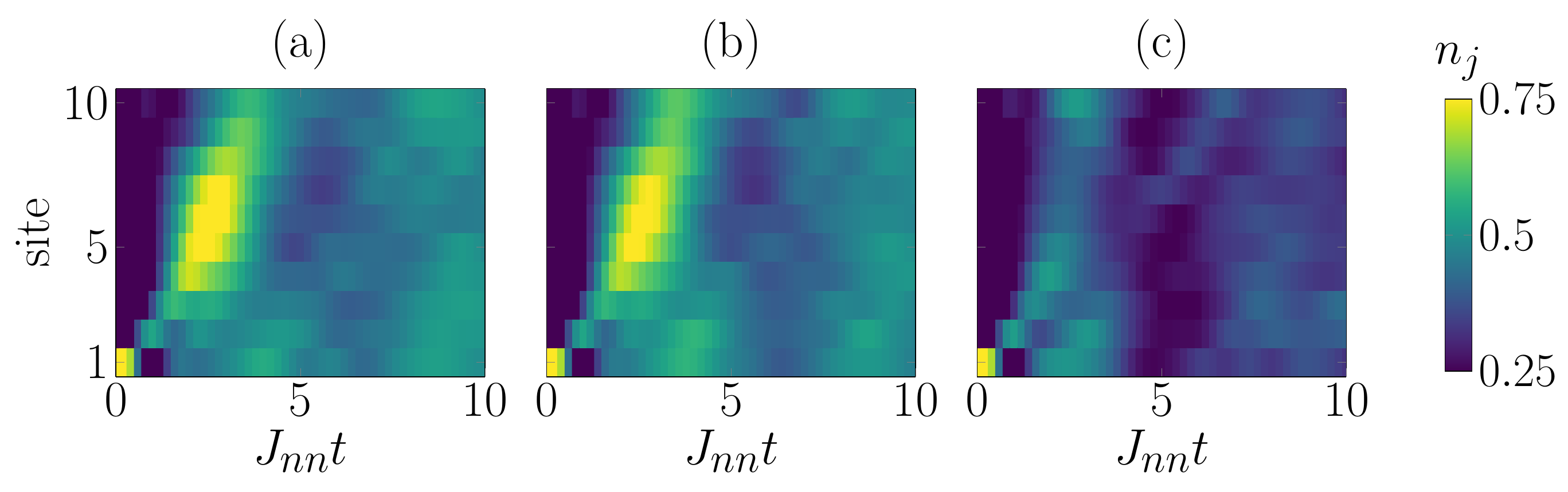}
\caption{Dynamics of the number of excitations for detunings $\Delta=0$ [panel (a)], $\Delta=0.1$ [panel (b)], and $\Delta=0.5$ [panel (c)]. Here we set $J_{nn}=\Omega=1$, $N=10$, $\theta=2\pi/5$ ($\ell=2$), and zero dephasing ($\gamma=0$).}
\label{fig:detunings}
\end{figure}

In Fig.~\ref{fig:detunings} we report the effect of detuning on the dynamics of the number of excitations. We consider the regime $J_{nn}/\Omega=1$ and two different values of detuning: $\Delta/J_{nn}=0.1$ and $\Delta/J_{nn}=0.5$. In the first case, the dynamics is similar to the one in the absence of detuning: the ring is crossed by a directional flow of excitations. When the detuning becomes comparable with the other energy scales of the system as the hopping and the Rabi frequency, clear differences emerge in the dynamics. The competition between Rabi and detuning comes into play: at intermediate times, the tilted yellow blob that indicates the flow of a huge amount of excitations disappears. Thus, the presence of a non-zero detuning has effects on the excitations flow. However, small values of detuning do not qualitatively alter the dynamics, maintaining the flow almost unchanged. 

\section{Excitation imbalance in the presence of dephasing}
\label{appC}

The dynamics of the system can be affected by the presence of dephasing. We have shown that the presence of small dephasing with respect to the typical energy scales of the system brings the populations $n_j$ to converge to $1/2$ at sufficiently long times (see Figs.~\ref{fig:N8_dynamics} and \ref{fig:N3_4_deph}). On the other hand, in the absence of dephasing, $n_j$ remains inhomogeneous on the ring at the times inspected, giving a non-zero excitation imbalance between different zones of the ring. In addition, the presence of an excitation imbalance can be a signal of directionality in the system since in the absence of phase the transport is symmetric in the chain.

To detect directionality via imbalance and study with it the effect of dephasing on the chiral flow of excitation, we introduce the following protocol: we consider a ring with an odd number of atoms ($N=9$) and  choose the initial state with a single excitation localized in the site $j=1$. We define $N_R=\sum_{j=2}^{(N+1)/2}n_j$, $N_L=\sum_{j=(N+1)/2+1}^{N}n_j$, $N_{exc}=\sum_{j=1}^{N}n_j$ and we introduce the excitation imbalance 
\begin{equation}
    IMB = \dfrac{N_R-N_L}{N_{exc}}.
\end{equation}
If the transport is symmetric with respect to $j=1$ or the number of excitations is homogeneous in the whole chain we expect $IMB=0$; an oscillation of the imbalance between positive and negative values can signal the presence of directionality in the system, the excitations move regularly from one side to the other of the ring.
\begin{figure*}[t]
\centering
\includegraphics[width=0.8\textwidth]{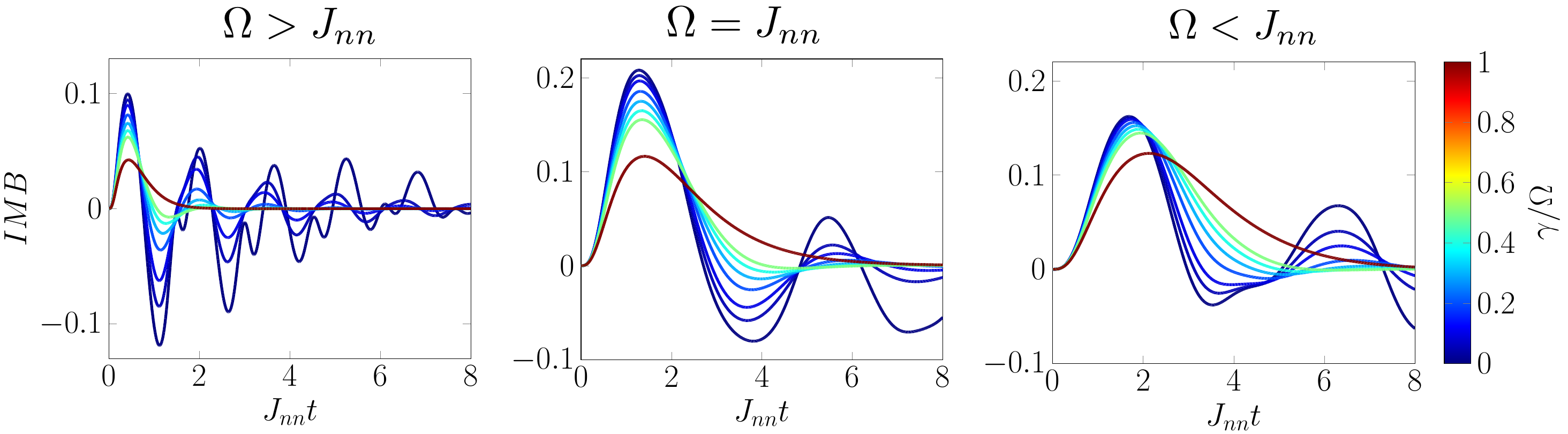}
\caption{Dynamics of the excitation imbalance in a ring composed by $N=9$ atoms and phase winding $\ell=2$ ($\theta=4\pi/9$). The different colors of the curves are different values of dephasing. Left figure: $J_{nn}=1$, $\Omega=4$. Central figure: $J_{nn}=\Omega=1$. Right figure: $J_{nn}=1$, $\Omega=0.5$. The dephasing is considered in the interval $\gamma/\Omega\in [0,1]$.}
\label{fig:imbalance}
\end{figure*}

In Fig.~\ref{fig:imbalance} we show the effect of dephasing on the imbalance dynamics in the three regimes analyzed in the main text ($\Omega > J_{nn}$, $\Omega = J_{nn}$, $\Omega < J_{nn}$). In all the three regimes the imbalance is positive at small times with a peaked behaviour in time, then it becomes negative and goes to zero oscillating in time. As expected, the oscillations are going to be suppressed increasing the dephasing strength indicating that the system is going to lose directionality. In the $\gamma/ \Omega \ll 1$ limit the dephasing does not affect too much the dynamics, the oscillations persist but they are suppressed in amplitude. Increasing $\gamma/ \Omega$ the suppression of the oscillations is more evident. In the $\gamma/ \Omega = 1$ limit the imbalance does not become negative in all the three regimes indicating that the majority of excitations is never in the right part of the ring. Thus, the flow of excitations along the ring is suppressed.
This result indicates that a small value of dephasing compared to the energy scales of the system is not dangerous for the flow, when the dephasing rate becomes big and comparable with the energy scales, the flow tends to be destroyed.

\section{Excitation position and velocity dynamics}\label{app:EXCPOS}
\begin{figure}[H]
\centering
\includegraphics[width=\columnwidth]{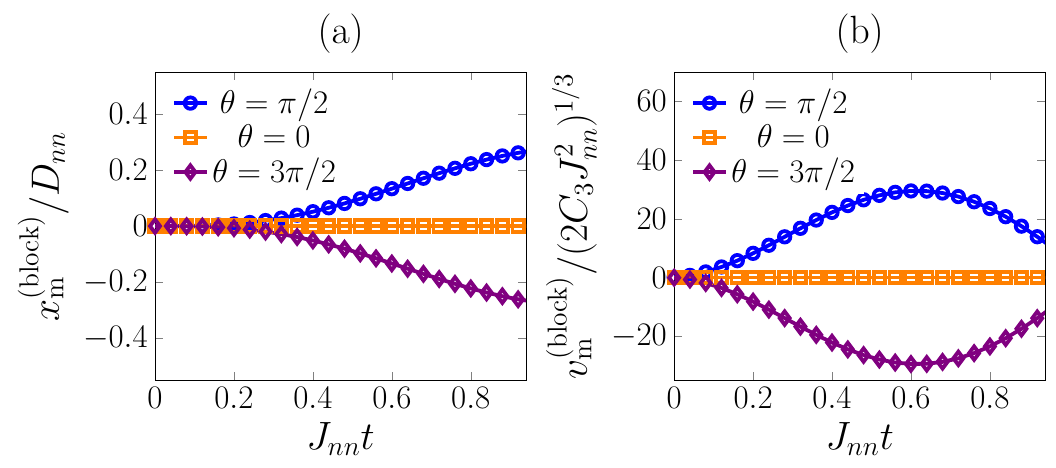}
\caption{Time behavior of $x_{\rm m}^{(\rm block)}/D_{nn}$ (a) and dimensionless velocity (b) in a system of $N=8$ atoms for three different values of the phase: $\theta=0$, $\theta=\pi/2$ ($\ell=2$) and its opposite $\theta=3\pi/2$ ($\ell=6$). $\Omega/J_{nn}=1$.}
\label{fig:XVt_theta_}
\end{figure}
Here we show briefly the dynamics of excitations mean position and its velocity in a block composed by 3 atoms. The two quantities are computed as in Sec.\ref{sec:Velocity}. 

The chiral character of the current in the dynamics of the center of mass is shown in Fig.~\ref{fig:XVt_theta_}. Initializing the excitation on the $j=0$ atom, for $\theta=0$ it travels to sites $j=-1$ and $j=+1$ with equal probability,  thus $x_{\rm m}^{(\rm block)} = v_{\rm m}^{(\rm block)} = 0$ for all times (see the orange curves in Fig.~\ref{fig:XVt_theta_}).
The presence of a nonzero phase originates a finite value of $x_{\rm m}^{(\rm block)}$. For $\theta=\pi/2$ the mean excitation position reaches a value $x_{\rm m}^{(\rm block)}/D_{nn}\neq 0$, i.e. the positive direction towards $j=1$ is preferred. The velocity shows a characteristic peak in time (see also Fig~\ref{fig:CM_main}(a)). For $\theta=3\pi/2$ (corresponding to $\theta=-\pi/2$), we find the same behavior in the opposite direction. Thus, our definition of center of mass can be used as a direct and easily observable measure of chirality for Rydberg excitations.

\section{Ground state nearest neighbors current}\label{app:GScurr}
We now focus our attention on the ground-state properties of the system. In Rydberg atoms systems the ground state can be prepared and studied using adiabatic protocols. For instance, the ground state properties of the Hamiltonian~\eqref{eq:int} in the isotropic limit have been recently studied in~\cite{chen2023continuous}. Here we study the ground state properties of the excitation current in a system whose dynamics is generated by an Hamiltonian like~\eqref{eq:Ham_rotated}. 

\begin{figure}[!t]
\centering
\includegraphics[width=\columnwidth]{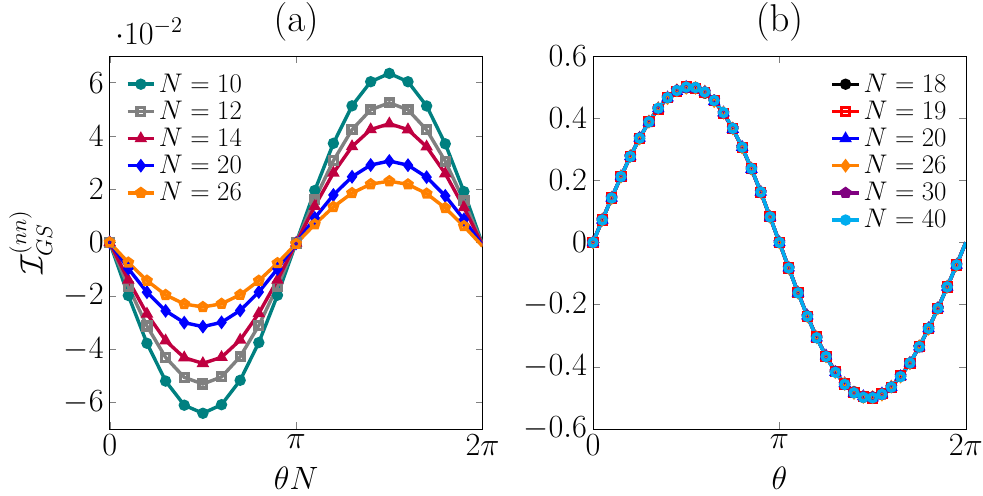}
\caption{Nearest-neighbor ground-state current for $\Omega=0.1$ as a function of $\theta N$ (a) and for $\Omega=8$ as a function of $\theta$ (b), The currents are compared for different values of the number of atoms. The nearest-neighbor hopping strength is $J_{nn}=1$.}
\label{fig:I_om}
\end{figure}

Due to the quasi-polynomial decrease of the interactions with the distance, to study properties of the ground-state current, it is important to focus on the dominant nearest-neighbor current ${\cal I}_j^{(nn)}$ of Eq.~\eqref{eq:currentNN}.
We define the corresponding average quantity as 
\begin{equation}
\mathcal{I}^{(nn)}=\dfrac{1}{N}\sum_{j}\mathcal{I}_{j}^{(nn)}  \end{equation}
and compute it for the ground state, obtaining $\mathcal{I}^{(nn)}_{GS}$. Long range currents are discussed in Appendix \ref{app:LR_currents}.

In Fig.~\ref{fig:I_om} we study how the current behaves for different values of the number of atoms in two different $\Omega / J_{nn}$
regimes: $\Omega \ll J_{nn}$ or $\Omega \gg J_{nn}$. We observe that, depending on the regime, the current presents different periods and different dependencies on the number of atoms. When $\Omega / J_{nn}  \ll 1$ the current has a period of $2\pi/N$ in $\theta$. For this reason, in Fig.~\ref{fig:I_om}(a) we report the current in the interval $\theta N \in [0,2\pi]$. Its behaviour is sinusoidal following $\mathcal{I}_{GS}^{(nn)}\sim -\sin(\theta N)$; its amplitude depends on the number of atoms, increasing the size of the system, the current decreases. In contrast, for larger $\Omega$, the period changes to $2\pi$. Indeed, when $\Omega / J_{nn} \gg 1$ (Fig.\ref{fig:I_om}(b)), the current has a different sinusoidal shape $\mathcal{I}_{GS}^{(nn)}\sim \sin(\theta)$. Moreover, the current does not show dependence on the number of atoms.

\section{Long range currents}\label{app:LR_currents}
The dynamics of the number of excitations is also influenced by long-range currents (i.e., beyond nearest neighbors), which also contribute to the ground-state current. However, due to the $1/D_{jk}^{3}$ dependence, their role is less important than that of the nearest-neighbor current. 
More precisely, we define the $r$-distance current 
\begin{equation}
\mathcal{I}_{j}^{(r)}=-i\dfrac{2C_{3}}{D_{r}}\left(e^{i r\theta}\sigma_{j}^{+}\sigma_{j+r}^{-}-e^{-i r\theta}\sigma_{j}^{-}\sigma_{j+r}^{+} \right)
\end{equation}
where $D_{r=|j-k|}\equiv D_r$. It is the excitation current between two atoms at a distance $r$. Using this notation, the nearest-neighbor current $\mathcal{I}^{(nn)}_{j}$ can be equivalently written as  $\mathcal{I}^{(1)}_{j}$. 

\begin{figure}[H]
\centering
\includegraphics[width=\columnwidth]{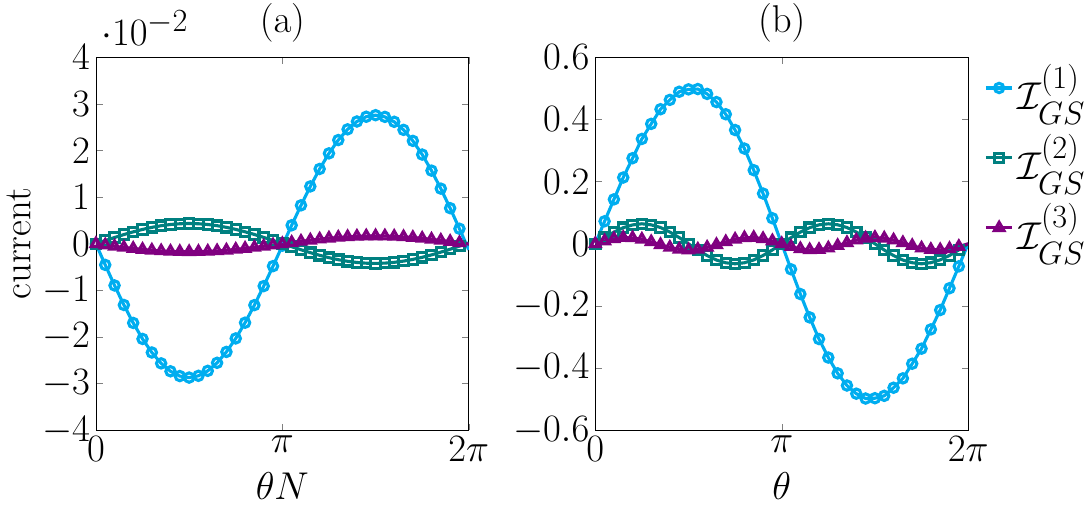}
\caption{(a) comparison between $\mathcal{I}^{(1)}$, $\mathcal{I}^{(2)}$ and $\mathcal{I}^{(3)}$ computed in the ground state for the $\theta\in [0,2\pi/N]$ and $\Omega=0.1$. (b) comparison between $\mathcal{I}^{(1)}$, $\mathcal{I}^{(2)}$ and $\mathcal{I}^{(3)}$ computed in the ground state in the $\theta\in [0,2\pi]$ domain for $\Omega=8$. We have $J_{nn}=1$ and $N=22$.}
\label{fig:IGS123}
\end{figure}

\begin{figure}[H]
\centering
\includegraphics[width=\columnwidth]{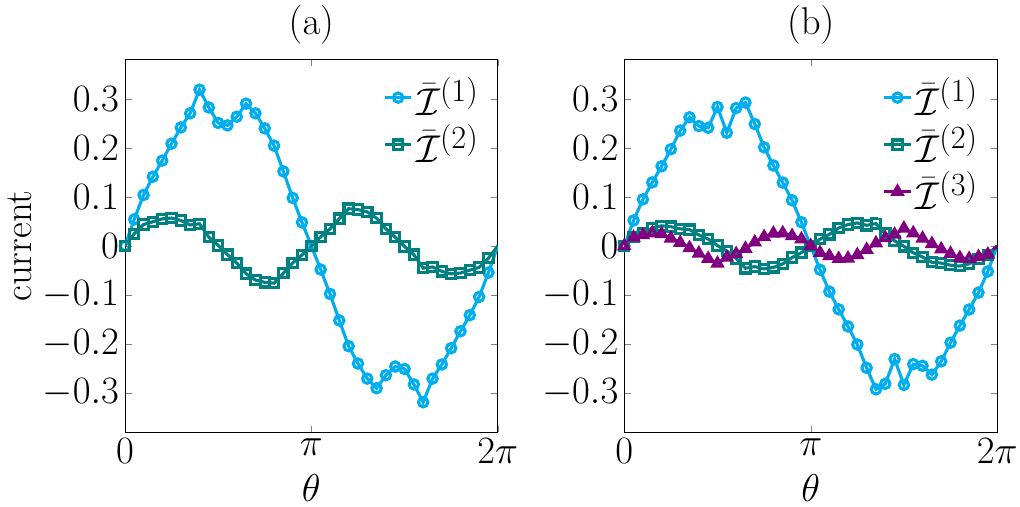}
\caption{(a) Comparison between the long-time-averaged currents $\bar{\mathcal{I}}^{(1)}$, $\bar{\mathcal{I}}^{(2)}$ as function of the phase for $N=5$ atoms. (b) Comparison between the long time averaged currents $\bar{\mathcal{I}}^{(1)}$, $\bar{\mathcal{I}}^{(2)}$ and $\bar{\mathcal{I}}^{(3)}$ for $N=6$ atoms. We fix $J_{nn}=1$ and $\Omega=8$.}
\label{fig:I123_LT}
\end{figure}

We compare the $3$-distance ground-state current $\mathcal{I}_{GS}^{(3)}$ with $\mathcal{I}_{GS}^{(1)}$ and $\mathcal{I}_{GS}^{(2)}$, in  Fig.~\ref{fig:IGS123}. We fix $J_{nn}=1$. For $\Omega=0.1$, the three currents have the same periodicity, although they differ in amplitude: $\mathcal{I}^{(r)}_{GS}$ decreases with $r$. Furthermore, the parity of $r$ affects the sign of the current: for $r=1,3$ ($r$ odd) the current behaves as $\mathcal{I}^{(r)}_{GS}\sim -\sin(\theta N)$, while for $r=2$ ($r$ even) it has opposite sign $\mathcal{I}^{(r)}_{GS}\sim \sin(\theta N)$. For $\Omega=8$,  the value of $r$ affects the periodicity of the current, with a period $2\pi/r$. The shape of the current follows $\mathcal{I}^{(r)}_{GS}\sim \sin(r\theta)$. Similar to $\Omega\ll1$, the amplitude of the current decreases with increasing $r$.
In Fig.~\ref{fig:I123_LT}, we analyze the long-time-averaged current for the three different $r=1,2,3$. We follow the same protocol proposed in Section \ref{sec:current}: we initialize the system in $\ket{\psi_{sup}}=\tfrac{1}{\sqrt{N}} \sum_j \ket{ \downarrow_1 \cdots \uparrow_j \cdots \downarrow_N}$, compute long time averages as in Eq.~\eqref{eq:Long_time_average}, and compare them with those of the ground state. In Fig.~\ref{fig:I123_LT} we report the long time-averaged currents for $N=5, 6$ with $\Omega/J_{nn}=8$. The shape of the currents is close to a sinusoidal, but not perfectly as for the ground state. Moreover, all $\bar{\mathcal{I}}^{(r)}$ have the same sign and period as for the ground state. The next to nearest neighbors current $\bar{\mathcal{I}}^{(2)}$ slightly differs from $\pi$. The shape of the current in the interval $[0,\pi/2]$ is the opposite to the one in $[3\pi /2,2\pi]$, as well as $[\pi/2,\pi]$ is opposite to $[\pi,3\pi/2]$. Overall, the behaviour in $[0,\pi]$ is mirrored but with opposite sign than in $[\pi,2\pi]$ due to the chiral nature of the flow. We find the same for $\bar{\mathcal{I}}^{(3)}$ in the interval $[0,2\pi /3]$. Despite the lack of periodicity, the long-time-averaged currents show chirality with $\bar{\mathcal{I}}^{(r)}(\theta)=-\bar{\mathcal{I}}^{(r)}(-\theta)$. Thus, they can reproduce the fundamental properties of the ground-state current.

\section{Quantum gates to measure the current of Rydberg excitations}\label{App:CurrentMeas}
The current operator of Rydberg excitations, as defined in the main text,
\begin{equation} {\mathcal{I}}_{j}^{(nn)} = i J_{nn}  \big( e^{i\theta} \sigma_{j}^{+} \sigma_{j+1}^{-} - {\rm H.c.} \big),
  \label{eq:currentNN_sup}
\end{equation}
can be rewritten as 
\begin{align} {\mathcal{I}}_{j}^{(nn)} =& \frac{1}{2} J_{nn} \Big[ - (\sin \theta) (\sigma_{j}^{x} \sigma_{j+1}^{x} + \sigma_{j}^{y} \sigma_{j+1}^{y}) + \nonumber \\& (\cos \theta)(\sigma_{j}^{x} \sigma_{j+1}^{y} - \sigma_{j}^{y} \sigma_{j+1}^{x}) \Big] \,,\label{eq:currentNNPauli_sup}
\end{align}
being a linear combination of four strings of Pauli operators. 
One can thus measure each of the Pauli strings individually. This is done by transforming into the eigenbasis of the respective operators with a unitary transformation $U\ket{\psi}$. For example, the term $\sigma_{j}^{y} \sigma_{j+1}^{x}$ is measured by transforming the $j$th atom into the $y$-basis via the unitary transformation $U_y= H S^\dagger$ ($H$ and $S$ being Hadamard and phase gate)~\cite{nielsen2002quantum}, and the $(j+1)$th atom into the $x$-basis with the transformation $U_x=H$. 
Thus, the expectation value is simply the probability of measuring the eigenvector multiplied with its eigenvalue. In our example, we have $\langle \sigma_{j}^{y} \sigma_{j+1}^{x}\rangle=P_{\uparrow\uparrow}+P_{\downarrow\downarrow}-P_{\downarrow\uparrow}-P_{\uparrow\downarrow}$, where $P_{ab}$ is the probability of measuring outcome $a\in\{\uparrow,\downarrow\}$, $b\in\{\uparrow,\downarrow\}$ on atom $j$ and $j+1$ respectively.

\begin{figure}[H]
\centering
\mbox{
\subfigure{\includegraphics[scale=1]{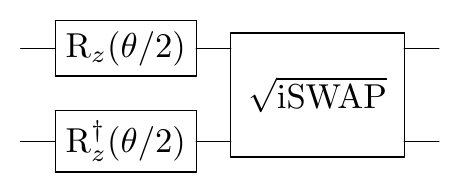}}
}
\caption{Unitary transformation into the eigenbasis of the current operator ${\mathcal{I}}_{j}^{(nn)}$ as function of the phase $\theta$.}
\label{fig:current_basiss}
\end{figure}

Alternatively, instead of measuring in four different Pauli basis, we can also directly compute the expectation of the current by transforming atoms $j$ and $j+1$ into the eigenbasis of the current operator ${\mathcal{I}}_{j}^{(nn)}$. In particular, we apply, on atoms $j$ and $j+1$, the unitary transformation $U_\text{d}$ that diagonalizes ${\mathcal{I}}_{j}^{(nn)}$.
Such transformation $U_\text{d}$ as function of $\theta$ is shown in Fig.~\ref{fig:current_basiss}, where $R_z(\theta)=\exp(- \tfrac12 i\theta \sigma^z)$ is a single-qubit rotation around the $z$-axis, and $\sqrt{\text{iSWAP}}=\exp \big[ \tfrac14 i\pi(\sigma^+_j \sigma^-_{j+1}+ \sigma^-_j \sigma^+_{j+1}) \big]$ is the square-root iSWAP gate~\cite{google2020hartree}.
The expectation value of the current is then given by $\langle{\mathcal{I}}_{j}^{(nn)}\rangle=J_{nn}(P_{\uparrow\downarrow}-P_{\downarrow\uparrow})$.

%

\end{document}